# Software-Defined Networking for Data Centre Network Management: A Survey

Jonathan Sherwin, Department of Computer Science,
Munster Techological University, Cork, Ireland[1]
Cormac J. Sreenan, School of Computer Science and Information Technology,
University College Cork, Cork, Ireland

*Abstract*

*Data centres are growing in numbers and size, and their networks expanding to carry larger amounts of traffic. The traffic profile is constantly varying, particularly in cloud data centres where tenants arrive, leave, and may change their resource requirements in between, and so the network configuration must change at a commensurate rate. Software-Defined Networking - programmatic control of network configuration - has been critical to meeting the demands of modern data centre network management, and has been the subject of intense focus by the research community, working in conjunction with industry. In this survey, we review Software-Defined Networking research targeting the management and operation of data centre networks.*

## 1  Introduction

IP traffic within data centres is expected to reach 14.7 ZettaBytes in 2021, with traffic between data centres coming to 2.8 ZettaBytes, according to the *Cisco Global Cloud Index 2017*. Traditional data centre traffic is declining, and cloud data centre traffic now accounts for up to 95% of the total traffic. To carry this amount of data, the cloud data centres themselves have grown and multiplied, compute power is being packed more densely, storage capacities are expanding, and data centre networks (DCNs) are growing as more devices are being connected at higher bandwidths. Business applications continue to move from private to public cloud, often to virtual data centres, and the resources underlying those virtual data centres may be distributed across physical data centres.

Data centre managers – and equally tenants, some of whom may themselves be managing virtual data centres – require sophisticated tools to abstract away the ever-increasing complexity, providing high-level control while allowing the critical management information to come through. Figure 1 illustrates some of the complexity of the situation, although many details and variations are omitted – e.g. an organisation may be a tenant of both physical data centres and virtual data centres; an operator of one data centre may be a tenant in another data centre; the management tools used in one data centre could be very different to the tools used in another data centre and might not lend themselves to being aggregated for control by multi-data centre management tools.

---

[1] *Contact Author* – jonathan.sherwin@cit.ie

In this paper, we describe the impact of Software-Defined Networking (SDN) in DCNs. SDN is the application of programmatic control over the resources and functions of a network to make the network more dynamically configurable to match the requirements of users and applications. SDN is often implemented using the open, standard OpenFlow protocol, although there are many other implementation options. SDN provides an attractive alternative to the traditional method of statically configuring network devices. A surge of networking research and development has been enabled by the rise of SDN, and in particular by the implementation of the OpenFlow protocol by network switch vendors.

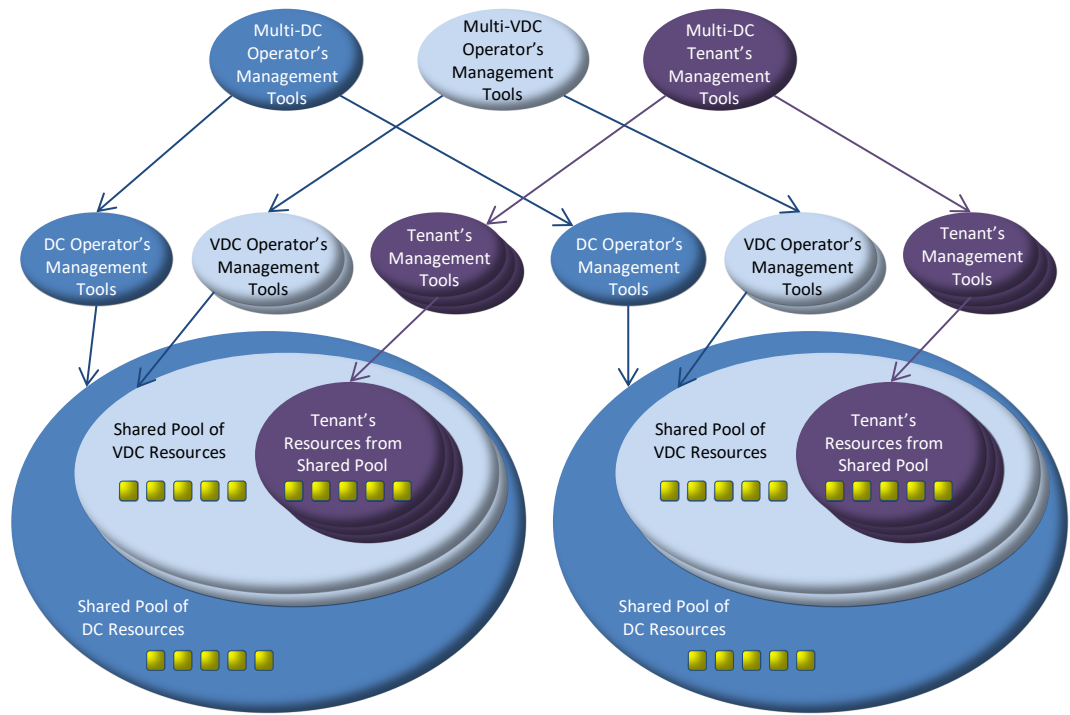

Figure 1: Cloud Multi-Data Centre with Virtual Network Operator

Data centres come in different levels of scale, from small server rooms to cloud-scale data centres. SDN has been most readily adopted in larger data centres where it promises efficient use of finite network resources in an environment of constant change – changing traffic patterns, changing applications, changing users of those applications, and even changing providers of the applications. We expect that the benefits of SDN will lead to its adoption in smaller, less dynamic data centres over time also.

SDN is sometimes conflated with network virtualisation. While related, they are different concepts. Network virtualisation abstracts the functionality of a physical network to create one or more virtual networks. The purpose could be to provide a simplified view of the underlying resources, or for controlled sharing of access to those resources between untrusted clients. SDN can be used to implement network virtualisation. Equally, SDN could be implemented over a virtualised network.

In the next sections, we describe the role of SDN in DCNs, focusing on uses that are of interest to data centre managers. For an earlier, wider-ranging survey of SDN we refer the reader to [1]. For an informative survey of network virtualisation for DCNs, see [2].

## 2    Role for SDN in data centres

For our review of SDN in DCNs we use a list of categories of concern to data centre managers, shown in Figure 2. Researchers have applied SDN to these different aspects of DCN operation. Key aspects of their work are described below.

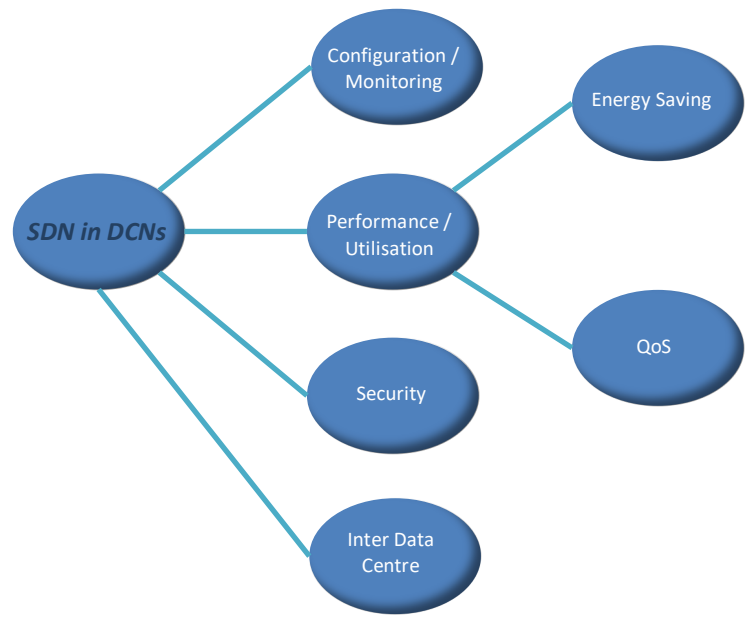

**Figure 2: SDN - Categories of Concern to Data Centre Managers**

## 2.1  Network Configuration / Monitoring

By configuration, we mean controlling the operation of the devices that make up a network. In a DCN, these are switches, routers, and middleboxes, all of which could be physical, but are increasingly also virtualised. A cloud-scale DCN could easily have more than 1,000 physical switches, and those switches may have different feature sets and capabilities.

Network management tools provide a high-level, abstracted view of the network to operators. The tools map high-level operator actions to lower-level interfaces provided by individual devices or software elements. Integrated into a suite of network management tools, SDN facilitates a centralised, unified view of network resources, and a single point for monitoring and configuration, allowing fine-grained control over how those resources are shared amongst applications and users. A network needs not be a collection of devices that are individually statically configured according to loose parameters of expected traffic levels and mixes, and that have limited capacity to dynamically react to changes in traffic conditions. SDN-based tools can allow the network manager to set a coherent policy, which is then implemented and policed automatically [3]. The policy can be complex and dynamic. In a cloud data centre, for example, the policy may need continuous adjustment as new tenants are added, old tenants leave, or existing tenants modify their service requirements. In fact, tenants should have the same control over their subset of DCN resources as the network operator has over the DCN as a whole, and SDN facilitates this.

Presenting a unified view of network resources is not trivial. Not only are there devices with different roles such as switch, router, middlebox to be integrated into that view, but devices performing the same role can have different capabilities – switches that have differently sized CAM tables (used storing OpenFlow rules and/or caching other forwarding information) [4]; switches that support different versions of OpenFlow; switches that can run user-defined code locally, and those that don't. Researchers have tackled some of these issues, and taken a software-defined approach in order to mask the underlying complexity with intelligent programs.

Configured network devices need to be monitored to ensure correct operation. There is a benefit to applying SDN here also – for example, by dynamically adjusting the amount of switch resources (CPU and CAM table) required for monitoring in an OpenFlow network, balancing the resource usage with the accuracy of measurement required [5]. The granularity of the information being collected, and the frequency at which it is collected, can be varied depending on requirements - at a congestion hotspot, it might be acceptable to reduce the detail of monitoring, to prioritise getting data packets through. On the other hand, more detailed monitoring might be required in some part of the network to positively identify a denial-of-service attack. SDN allows this flexibility in contrast to established practice of dedicating a fixed set of resources to monitoring in a network. Recently, attention has focussed on network telemetry – collecting data on specific packets sent through the data plane for the purpose of fault detection and troubleshooting [6].

## 2.2 Performance / Utilisation

In the largest data centres, for example those owned and run by public cloud providers, a small increase in performance (greater throughput, lower latency) can translate to a competitive advantage. Techniques used to improve performance include network traffic splitting and traffic engineering, layer 4-7 load balancing, and virtual-machine (VM) consolidation.

Network traffic splitting makes use of redundant network paths within a network. Redundant paths are part of a network design, allowing for device or link failure; however, they represent a significant financial investment, so the motivation is to make use of the resources during normal network operation thereby increasing overall utilisation of network resources. ECMP is a non-SDN technique commonly used in data centres that splits traffic, on a per-flow basis, across available paths using layer 2 or layer 3 devices. However, ECMP does not take the load of each individual flow into account, so traffic is not fully balanced. An SDN-based solution can do better than simple traffic splitting: Planck [7] builds on previous work to monitor data centre switches and provides an application that performs fine-grained traffic engineering by re-routing congested flows within milliseconds (<10ms on 1Gb/s and 10Gb/s networks).

Layer 4 load balancing is the distribution of client flows across several data centre servers providing a common service identified by TCP port number. Rather than replace a heavily loaded server with a bigger server – 'scale-up' – load balancing allows the problem to be solved by adding more servers to share the work – 'scale-out' – and this is transparent to clients. Layer 4 load balancing has traditionally been implemented using dedicated hardware appliances, although more recently vendors have offered virtual appliances. Duet [8] is a combined software and hardware SDN load balancer. Most of the load balancing work is offloaded to DCN switches, tapping under-used resources in switches already in place in the data centre network. This naturally scales with the data centre with minimum cost. Software load-balancing in Duet primarily provides fault-tolerance. Duet uses vendor-specific APIs for controller-to-switch communication, rather than OpenFlow.

Layer 7 traffic steering directs packets to servers based on deep packet inspection (DPI) of application-layer data – for functions such as TCP splicing, NAT, layer 7 server selection, or firewall. While traffic steering can be done through other methods – e.g., MPLS tunnels – using SDN allows the steering to be dynamically reconfigured in response to changes in policy, load or other factors. The authors of [9] propose an SDN architecture in which application-specific packet processing capabilities are distributed to switches in a data centre as 'apps', eliminating the need for dedicated appliances to act as middleboxes. The apps on an individual switch can be chained together to achieve service-chaining, simplifying the path along which packets must be steered.

One way to reduce the latency between two communicating VMs within a data centre is to move them 'closer' to each other – shortening the path between them. This also reduces overall network utilisation because less links and intermediate devices are needed to carry the inter-VM traffic. The role of SDN is firstly to ensure that data packets continue to reach the VM after it has been migrated, so that the move is transparent to the VM itself and communicating partners – in particular so that no TCP connections are dropped. Secondly, SDN can be used to reserve or allocate network resources to facilitate a quick migration of a VM across the DCN. Of course, the benefit of re-locating a VM must outweigh the cost of migration which itself uses network, CPU, and storage resources.

Installation of flow-rules in hardware switch CAM tables is relatively slow, both for adding a single rule and the rate at which multiple rules can be added. Especially a multi-tenant DCN, the number of flows is large and has a high degree of churn – so the rate at which rules can be added will be a limiting factor on the DCNs ability to carry new flows. One approach tackles the issue by temporarily directing packets for a new flow across hardware switches through preconfigured tunnels, while waiting for the hardware switches to install the rules required to forward packets of the flow natively, at which point remaining packets of the flow (if still incomplete) can be redirected across the non-tunnelled path [10].

All switches are not created equal – some may have larger flow-tables, higher speed interfaces, faster CPUs. Researchers have addressed the problem of where switches should be placed in a DCN to take best advantage of their individual flow-table capacities [11].

### 2.2.1 Quality of Service

Quality of Service (QoS) implies giving some traffic classes higher priority service than others, possibly to the extent of giving guarantees on latency, bandwidth, and/or packet-loss to specific data flows.

In a data centre, congestion hotspots may be minimised firstly by generous provisioning of bandwidth, and secondly by careful planning and application of traffic engineering to make balanced use of bandwidth, or to deliberately skew the balance in favour of specific, recurring flows. However, lack of bandwidth is only one of the causes of congestion. Other factors that contribute are buffer capacity in switches and routers, and operations that must be applied on a per-packet basis to packets received or forwarded. In addition, in a cloud data centre, traffic profiles change as tenants arrive and leave, scale their usage up and down, and change the applications they are using – a scenario that is too dynamic to suit traditional traffic engineering. QoS has a role in isolating traffic that requires guaranteed service from the effects of unavoidable congestion, and in prioritising resource allocation to other traffic according to administrative policy. QoS implemented through SDN can flexibly accommodate different service granularity for different flows, and constantly changing policies. SDN can be used to dynamically re-route flows out of a congestion hotspot while maintaining service guarantees. DCN applications that benefits from service guarantees includes distributed databases, map-reduce, and virtual networking [12].

One way that QoS has been implemented for Data Centres using SDN is to modify applications to request guarantees in advance for their traffic. In [13], requests are added to a policy tree which is used to resolve conflicts between competing demands – important in an oversubscribed network. The researchers in [12] implement explicit path control - allowing an application to choose paths through the network for its traffic in order to achieve bandwidth/latency guarantees.

### 2.2.2 Energy Saving

As data centres grow in size and number, the amount of energy consumed rises. This has motivated researchers and industry to control and minimise data centre energy usage. Initially, the focus was on consolidating virtual machine in order to power down servers during off-peak times, but researchers' attention has widened to consider the network as it consumes a significant portion of a data centre's total power usage. A data centre network in particular is usually designed to meet peak demand, with generous amounts of bandwidth available to avoid bottlenecks anywhere in the network, and redundant links and devices to minimise the likelihood of the network becoming partitioned due to failure. These requirements incur an up-front equipment cost, and also an on-going cost due to all links and devices being powered constantly although the network operates at a fraction of its full capacity for most of the time.

SDN can help with the aim of reducing energy consumption, by allowing flows to be routed (or re-routed) through a subset of the switches and links in a DCN. The switches and links that are not in use can be powered down or put in an energy-saving standby state until required for use later, perhaps when demand peaks. A simple approach [14] uses Energy-aware Traffic Engineering (ETE) to disable inter-switch links when observed traffic patterns suggest they are not required (e.g. at night). For a more complex approach, in [15] the power-profile of the switches can be considered, and the more power-hungry devices powered down first when there is a choice; for the switches that are kept active, the traffic distribution can be designed to target specific utilisation levels on individual switches, with the switch CPU clock and individual port speeds set to match the load to eke out better power savings. SDN enables these approaches firstly through the network-wide view it provides, allowing collection of data on what flows are currently in the network. Secondly, all switches can be reconfigured through the control plane to free up devices and links before powering them down – critical to avoid packet loss and latency spikes.

## 2.3 Security

SDN has many benefits in for network security. In the context of a DCN, SDN can be used to implement a security policy, in place of, or more likely augmenting, the traditional security mechanisms such as VLANs and firewalls. If the security policy changes, an SDN-based solution should dynamically change the operation of the network to match. If the security policy is too complex and nuanced to be realised entirely with static mechanisms or on devices with limited resources, SDN can provide a means to dynamically reconfigure both based on time-of-day, demand, traffic profile, or other criteria.

While the VLAN (or newer VXLAN, NVGRE and GENEVE) functionality of traffic isolation can be provided solely with SDN, it is interesting to see how SDN can leverage VLAN features to do more. For example, deciding VLAN membership based on arbitrary criteria (for example, tenant ID, as in [16]), or re-writing VLAN tags as packets cross a network in order to merge or partially overlap VLANs.

SDN is particularly useful for implementing security policy in a cloud data centre where tenants are allowed to do self-service provisioning – creating and migrating virtual machines, creating virtual networks, adding and configuring virtual security devices, setting their own security policies for their virtual cloud [17]. Validating and merging these sub-policies, and verifying that the result does not violate the provider's super-policy, is a complex task that can be solved with the help of SDN: for example, by translating the policies to underlay device configurations, and verifying implemented policy through active packet probing.

The widely-used technique of anomaly detection has benefited from integration in an SDN framework [18], to achieve the scalability required for operating machine-learning-based network anomaly detection in a cloud DCN.

## 2.4 Inter-Data Centre

Data centre owners quite often run more than one data centre – for redundancy, disaster recovery, or locating services close to groups of users (e.g., to minimise latency or Internet backbone bandwidth utilisation). For the same reasons, data centre tenants may have operations in multiple data centres, which need not all belong to the same provider.

The WAN links and Internet-based connections between data centres are different to the networks inside data centres. For example, the links within a provider's network are typically provisioned for low utilisation to give customers the illusion of high reliability – traffic is transparently rerouted in case of link or equipment failure. For a data centre operator who is also a WAN provider, there is an opportunity to trade reliability for lower cost – by using lower bandwidth links run at high utilisation. In case of link failure, fast action is required to minimise transient congestion, and SDN has been used to achieve this with low overhead [19].

In cloud multi-data centres, operators can give tenants the capability of creating, scaling up and down, and terminating virtual links between tenant networks in different data centres, across the physical links the operator already has in place [20]. SDN provides the mechanism for the changes to virtual links to be effected on a self-service basis, including specifying parameters such as bandwidth requirements and QoS options. Traditionally, such changes would have to be requested by the tenant through a ticketing system and implemented manually by the operator.

The mechanism for creating, modifying or tearing down virtual links can itself be put under programmatic control, for example to reserve resources required to migrate VMs from one data centre to another and to maintain network connectivity for VMs that have been so migrated [21]. The benefit of SDN here is being able to automate the configuration of the network connections between data centres with the very fine degree of control required when those connections are simultaneously carrying multiple tenants' traffic for which performance guarantees must be maintained.

## 3    Conclusion

We have identified key areas for the role of SDN in DCNs: configuration / monitoring, performance / utilisation (with sub-areas of energy saving and QoS), security, and inter-data centre operations. Within each of these areas we discussed the benefits brought by SDN and how those benefits have been realised through the work of the research community. It is clear that SDN provides critical benefits for DCN managers, and the area is the subject of continuous active research.

*References*

[1]    D. Kreutz, F. M. V. Ramos, P. E. Verissimo, C. E. Rothenberg, S. Azodolmolky, and S. Uhlig, "Software-defined networking: A comprehensive survey," *Proceedings of the IEEE,* vol. 103, no. 1, pp. 14-76, 2014, Art no. 6994333, doi: 10.1109/JPROC.2014.2371999.

[2]    M. F. Bari *et al.*, "Data center network virtualization: A survey," *IEEE Communications Surveys and Tutorials,* vol. 15, no. 2, pp. 909-928, 2013, Art no. 6308765.

[3]    K. Hyojoon and N. Feamster, "Improving network management with software defined networking," *Communications Magazine, IEEE,* vol. 51, no. 2, pp. 114-119, 2013, doi: 10.1109/MCOM.2013.6461195.

[4]    N. Kang, Z. Liu, J. Rexford, and D. Walker, "Optimizing the "One big switch" abstraction in software-defined networks," in *CoNEXT 2013 - Proceedings of the 2013 ACM International Conference on Emerging Networking Experiments and Technologies*, 2013, pp. 13-24. [Online]. Available: http://www.scopus.com/inward/record.url?eid=2-s2.0-84893398564&partnerID=40&md5=198adea3b54a99cfcb044ee2f143cc52

[5]    M. Moshref, M. Yu, R. Govindan, and A. Vahdat, "DREAM: Dynamic resource allocation for software-defined measurement," in *SIGCOMM 2014 - Proceedings of the 2014 ACM Conference on Special Interest Group on Data Communication*, 2014, pp. 419-430, doi: 10.1145/2619239.2626291. [Online]. Available: http://www.scopus.com/inward/record.url?eid=2-s2.0-84907377109&partnerID=40&md5=61a23e683e79618fbafb5aa01ad1f1c9

[6]     N. Foster, N. McKeown, J. Rexford, G. Parulkar, L. Peterson, and O. Sunay, "Using deep programmability to put network owners in control," *SIGCOMM Comput. Commun. Rev.,* vol. 50, no. 4, pp. 82–88, 2020, doi: 10.1145/3431832.3431842.

[7]     J. Rasley *et al.*, "Planck: Millisecond-scale monitoring and control for commodity networks," in *SIGCOMM 2014 - Proceedings of the 2014 ACM Conference on Special Interest Group on Data Communication*, 2014, pp. 407-418, doi: 10.1145/2619239.2626310. [Online]. Available: http://www.scopus.com/inward/record.url?eid=2-s2.0-84907359489&partnerID=40&md5=a253059141e76db2439c39c3a154c74c

[8]     R. Gandhi *et al.*, "Duet: Cloud scale load balancing with hardware and software," in *SIGCOMM 2014 - Proceedings of the 2014 ACM Conference on Special Interest Group on Data Communication*, 2014, pp. 27-38, doi: 10.1145/2619239.2626317. [Online]. Available: http://www.scopus.com/inward/record.url?eid=2-s2.0-84907321982&partnerID=40&md5=10a09583e46cf66edae37677e6b39594

[9]     H. Mekky, F. Hao, S. Mukherjee, Z. L. Zhang, and T. V. Lakshman, "Application-aware data plane processing in SDN," in *HotSDN 2014 - Proceedings of the ACM SIGCOMM 2014 Workshop on Hot Topics in Software Defined Networking*, 2014, pp. 13-18, doi: 10.1145/2620728.2620735. [Online]. Available: http://www.scopus.com/inward/record.url?eid=2-s2.0-84907009902&partnerID=40&md5=18d5f099f20d32fd240e1ae6048fed0d

[10]    J. Sherwin and C. J. Sreenan, "Reducing the latency of OpenFlow rule changes in data centre networks," in *2018 21st Conference on Innovation in Clouds, Internet and Networks and Workshops (ICIN)*, 19-22 Feb. 2018 2018, pp. 1-5, doi: 10.1109/ICIN.2018.8401629.

[11]    G. Shen, Q. Li, S. Ai, Y. Jiang, M. Xu, and X. Jia, "How Powerful Switches Should be Deployed: A Precise Estimation Based on Queuing Theory," in *IEEE INFOCOM 2019 - IEEE Conference on Computer Communications*, 29 April-2 May 2019 2019, pp. 811-819, doi: 10.1109/INFOCOM.2019.8737629.

[12]    S. Hu *et al.*, "Explicit path control in commodity data centers: design and applications," presented at the Proceedings of the 12th USENIX Conference on Networked Systems Design and Implementation, Oakland, CA, 2015.

[13]    A. D. Ferguson, A. Guha, C. Liang, R. Fonseca, and S. Krishnamurthi, "Participatory networking: An API for application control of SDNs," in *SIGCOMM 2013 - Proceedings of the ACM SIGCOMM 2013 Conference on Applications, Technologies, Architectures, and Protocols for Computer Communication*, 2013, pp. 327-338, doi: 10.1145/2486001.2486003. [Online]. Available: http://www.scopus.com/inward/record.url?eid=2-s2.0-84883299005&partnerID=40&md5=46e67c26a2f7f040faf692e5149a6032

[14]    J. Ba, Y. Wang, X. Zhong, S. Feng, X. Qiu, and S. Guo, "An SDN energy saving method based on topology switch and rerouting," in *NOMS 2018 - 2018 IEEE/IFIP Network Operations and Management Symposium*, 23-27 April 2018 2018, pp. 1-5, doi: 10.1109/NOMS.2018.8406202.

[15]    N. Tran Manh, T. Nguyen Huu, T. Ngo Quynh, H. Hoang Trung, and S. Covaci, "Energy-aware routing based on power profile of devices in data center networks using SDN," in *Electrical Engineering/Electronics, Computer, Telecommunications and Information Technology (ECTI-CON), 2015 12th International Conference on*, 24-27 June 2015 2015, pp. 1-6, doi: 10.1109/ECTICon.2015.7207042.

[16]    D. Arora, T. Benson, and J. Rexford, "ProActive routing in scalable data centers with PARIS," in *DCC 2014 - Proceedings of the ACM SIGCOMM 2014 Workshop on Distributed Cloud Computing*, 2014, pp. 5-10, doi: 10.1145/2627566.2627571. [Online]. Available: http://www.scopus.com/inward/record.url?eid=2-s2.0-84907327718&partnerID=40&md5=88f3628d70034e18431ff1924f7d144c

[17]    R. Cohen *et al.*, "An intent-based approach for network virtualization," in *Proceedings of the 2013 IFIP/IEEE International Symposium on Integrated Network Management, IM 2013*, 2013, pp. 42-50.

[Online]. Available: http://www.scopus.com/inward/record.url?eid=2-s2.0-84883484060&partnerID=40&md5=33835fe721726b09a32273e97e948440

[18]   S. Lee, J. Kim, S. Shin, P. Porras, and V. Yegneswaran, "Athena: A Framework for Scalable Anomaly Detection in Software-Defined Networks," in *2017 47th Annual IEEE/IFIP International Conference on Dependable Systems and Networks (DSN)*, 26-29 June 2017 2017, pp. 249-260, doi: 10.1109/DSN.2017.42.

[19]   J. Zheng, H. Xu, G. Chen, and H. Dai, "Minimizing transient congestion during network update in data centers," in *Proceedings - International Conference on Network Protocols, ICNP*, 2016, vol. 2016-March, pp. 1-10, doi: 10.1109/ICNP.2015.33. [Online]. Available: https://www.scopus.com/inward/record.uri?eid=2-s2.0-84969764481&partnerID=40&md5=2afe79d86aba027e657d12f01f56f626

[20]   S. Baucke, R. Ben Ali, J. Kempf, R. Mishra, F. Ferioli, and A. Carossino, "Cloud Atlas: A Software Defined Networking Abstraction for Cloud to WAN Virtual Networking," in *Cloud Computing (CLOUD), 2013 IEEE Sixth International Conference on*, June 28 2013-July 3 2013, pp. 895-902, doi: 10.1109/CLOUD.2013.44.

[21]   J. Liu, Y. Li, and D. Jin, "SDN-based live VM migration across datacenters," presented at the Proceedings of the 2014 ACM conference on SIGCOMM, Chicago, Illinois, USA, 2014.

BIOGRAPHIES

Jonathan Sherwin (jonathan.sherwin@cit.ie) received a B.Sc. degree in Computing from Cork Institute of Technology (CIT), Ireland in 1990, and an M.Sc. in Computer Science from University College Cork (UCC), Ireland in 2004. Following an internship in Apple, California in 1990, he worked as a Software Development Engineer and then Engineering Team Lead in Microsoft, Ireland until 1998. Since 1999, he has been a lecturer in the Department of Computing, CIT. He is also currently pursuing a PhD in UCC in the area of software-defined networking in data centre networks, supported by CIT. His other research interests include network management, wireless networks, internet of things, and media streaming. He is a member of the ACM.

Cormac J. Sreenan (cjs@cs.ucc.ie) received the Ph.D. degree in computer science from Cambridge University. He is a professor of computer science at University College Cork (UCC) in Ireland. Prior to joining UCC in 1999, he was on the research staff at AT&T Labs—Research, Florham Park, New Jersey, and at Bell Labs, Murray Hill, New Jersey. At Cork he directs the Mobile & Internet Systems Lab., which conducts research on the Internet of Things, SDN and video streaming. He is a PI and Deputy-Director for the CONNECT Research Centre funded by Science Foundation Ireland and numerous industry partners. He was elected a Fellow of the British Computer Society in 2005 and is a member of the IEEE and ACM.